\documentclass[conference]{IEEEtran}
\IEEEoverridecommandlockouts
\usepackage{cite}
\usepackage{amsmath,amssymb,amsfonts}
\usepackage{algorithmic}
\usepackage{graphicx}
\usepackage{textcomp}
\usepackage{hyperref}
\usepackage[table,xcdraw]{xcolor}
\usepackage{colortbl}
\usepackage{pifont}
\usepackage{tabularx}
\usepackage{cleveref}
\crefname{section}{section}{sections}
\Crefname{section}{Section}{Sections}
\crefname{table}{table}{tables}
\Crefname{table}{Table}{Tables}
\crefname{figure}{figure}{figures}
\Crefname{figure}{Figure}{Figures}
\def\BibTeX{{\rm B\kern-.05em{\sc i\kern-.025em b}\kern-.08em
    T\kern-.1667em\lower.7ex\hbox{E}\kern-.125emX}}
\usepackage{color}

\begin{document}

\title{Generative Large Language Model usage in Smart Contract Vulnerability Detection\\

\thanks{Generative AI, including ChatGPT and Cursor IDE, has been used to assist with code and latex table formatting.}
}

\author{
\IEEEauthorblockN{Peter Ince}
\IEEEauthorblockA{\textit{Faculty of IT} \\
\textit{Monash University}\\
Clayton, Australia \\
peter.ince1@monash.edu}
\and
\IEEEauthorblockN{Jiangshan Yu}
\IEEEauthorblockA{\textit{School of Computer Science} \\
\textit{University of Sydney}\\
Darlington, Australia \\
jiangshan.yu@sydney.edu.au}
\and
\IEEEauthorblockN{Joseph K. Liu}
\IEEEauthorblockA{\textit{Faculty of IT} \\
\textit{Monash University}\\
Clayton, Australia \\
joseph.liu@monash.edu}
\and
\IEEEauthorblockN{Xiaoning Du}
\IEEEauthorblockA{\textit{Faculty of IT} \\
\textit{Monash University}\\
Clayton, Australia \\
xiaoning.du@monash.edu}
}

\maketitle

\begin{abstract}
Recent years have seen an explosion of activity in  Generative AI, specifically Large Language Models (LLMs), revolutionising applications across various fields. Smart contract vulnerability detection is no exception; as smart contracts exist on public chains and can have billions of dollars transacted daily, continuous improvement in vulnerability detection is crucial. This has led to many researchers investigating the usage of generative large language models (LLMs) to aid in detecting vulnerabilities in smart contracts.

This paper presents a systematic review of the current LLM-based smart contract vulnerability detection tools, comparing them against traditional static and dynamic analysis tools Slither and Mythril. Our analysis highlights key areas where each performs better and shows that while these tools show promise, the LLM-based tools available for testing are not ready to replace more traditional tools. We conclude with recommendations on how LLMs are best used in the vulnerability detection process and offer insights for improving on the state-of-the-art via hybrid approaches and targeted pre-training of much smaller models.
\end{abstract}

\begin{IEEEkeywords}
Ethereum, Smart Contracts, Vulnerability Detection, Large Language Models, Evaluation
\end{IEEEkeywords}
\section{Introduction}
Smart contracts are essential components of decentralised ecosystems that run on blockchains, such as Ethereum\cite{buterin_ethereum_2014}, which enable applications such as Decentralised Finance (DeFi) and Decentralised Autonomous Organisations (DAOs). These contracts are often deployed to public blockchains (often with their verified source code published), and as they cannot be natively updated once deployed (although developers can use upgradeable smart contract patterns), ensuring their security is critical.

Traditional vulnerability detection tools, such as Slither\cite{feist_slither_2019} for static analysis and Mythril\cite{consensys_mythril_2023} for symbolic execution, have greatly improved smart contract security. However, they are not without limitations. Static analysis tools tend to be fast but often produce false positives. Dynamic analysis tools tend to produce fewer false positives but can be slow and computationally expensive. Also, static and dynamic analysis tools can struggle to detect nuanced logic vulnerabilities.

Since the release of ChatGPT in November 2022\cite{openai_introducing_2022}, Large Language Models (LLMs) have become an ever-increasing component of our lives and work.
While LLMs as a category include several approaches, the generative (or next token prediction) style has become synonymous with the term.

Generative LLMs have shown promise in diverse fields, such as Healthcare, Finance and Education\cite{hadi_survey_2023}.
Growth has also been seen in the use of LLMs for software security in areas such as fuzzing\cite{hu_augmenting_2023}, source code inspection\cite{szabo_new_2023}, automated program repair\cite{yang_automated_2024} and detecting illicit activity\cite{nicholls_enhancing_2023}.

Thus far, there have been many different approaches for utilising LLMs in various forms for blockchain security, including;
\begin{itemize}
    \item Training of a custom LLM from Ethereum transactions to DeFi contracts for detection of suspicious transactions in the mempool before they reach the contract\cite{gai_blockchain_2023}
    \item Detection and resolution of access control bugs in smart contracts\cite{zhang_acfix_2024}
    \item Efficient generation of vulnerability-free smart contract code\cite{storhaug_efficient_2023}
\end{itemize}

However, the most common use of LLMs in blockchain security is to detect vulnerabilities in smart contracts.
There are many approaches to incorporating, training, and evaluating LLMs (specifically generative LLMs) for detecting smart contract vulnerabilities. Yet, to the best of our knowledge, there has been no detailed study of the tools with an evaluation and discussion of their effectiveness.

In this paper, we conduct a comprehensive and detailed study of the current vulnerability detection tools that include LLMs as a primary component.
For this analysis, we evaluate how LLM(s) are used in the vulnerability detection process, the techniques that differentiate their tool from others, and the data they are trained on.
We then evaluate the available tools where possible and compare their ability to find vulnerabilities against Slither\cite{feist_slither_2019} and Mythril\cite{consensys_mythril_2023}. In addition, we compare the tools' speed, cost, and runtime.

\subsection{Our Contributions}
\begin{itemize}
    \item We present a comprehensive, up-to-date study on LLM usage focused on smart contract vulnerability detection,  providing a detailed comparison with traditional static and dynamic analysis tools like Slither and Mythril.
    \item We thoroughly evaluate open-source LLM-based tools, identifying their strengths and weaknesses across multiple vulnerability types. Our benchmarking provides critical insights into the capabilities of LLMs, revealing that while they perform well in detecting specific vulnerabilities, they are not yet ready to replace traditional tools.
    \item We identify the most effective approaches across all analysed tools and show the best performance comes from unique hybrid approaches (such as LLM4Fuzz\cite{shou_llm4fuzz_2024}) and the counter-intuitive approach of small models pre-trained on targeted data (\cite{zeng_solgpt_2024}).
\end{itemize}



\section{Background}
\subsection{Large Language Models}
Large Language Models (LLMs) are a form of artificial intelligence pre-trained on a large corpus of data. Although many organisations that train LLMs do not disclose the full dataset they are trained on, the data corpus is likely made up of several components;
\begin{itemize}
    \item Data scraped from the web and websites
    \item Code from open source code repositories (e.g. the StarCoder family of models\cite{li_starcoder_2023} where trained on The Stack\cite{Kocetkov2022TheStack})
    \item Data from existing open-source datasets
    \item Data from private datasets of books
    \item data from social networks/sites
\end{itemize}

The current generation of LLMs are primarily built using a Transformers-based architecture\cite{vaswani_attention_2017}. The transformers architecture has 3 main variants;
\begin{enumerate}
    \item Encoder only - ideal for tasks like classification. Models such as CodeBERT\cite{feng_codebert_2020} and BERT\cite{devlin_bert_2019} fall into this category.
    \item Encoder-decoder - ideal for tasks such as translation and summarisation as the input can be encoded to a vector, and the decoder can generate the output independently. Examples of this model type include BART\cite{lewis_bart_2019} and CodeT5\cite{wang_codet5_2021}.
    \item Decoder only - these models are great for text generation tasks, and their simplicity makes them easier to scale. Examples of this model type are OpenAI's GPT Series, GPT-2\cite{radford2019language}, GPT-3\cite{brown_language_2020} and GPT-4\cite{openai_gpt-4_2023}.
\end{enumerate}

\subsubsection{Generative Pre-trained Transformers}
Generative Pre-trained Transformers (GPTs) are models that use a decoder-only Transformer architecture and are pre-trained using unsupervised learning on large corpora's of data, and then further tuned on more specific fine-tuning on tasks\cite{radford_improving_2018}.

The decoder-only Transformer architecture and pre-training approach GPTs introduced became the basis for most of the generative LLMs we see today. This was then improved in InstructGPT\cite{ouyang_training_2022}, where they used user feedback to improve their models using the Reinforcement Learning from Human Feedback technique (RLHF)\cite{ouyang_training_2022}. 

\subsection{Retrieval-Augmented Generation}
Retrieval-Augmented Generation (RAG), is a process whereby a Large Language Model is used in conjunction with an external "memory", or knowledge-base, to achieve better results than with the language model alone\cite{lewis_retrieval-augmented_2020}. RAG can be used to supplement the existing knowledge base of an LLM as an alternative to fine-tuning.

\subsection{Smart Contract Vulnerability Detection}
Beyond the fiscal damages associated with smart contract exploits, it impacts the perception of trust in the blockchain ecosystem and limits the adoption of the technology on a wider scale.

There are two primary kinds of vulnerability detection tools;
\subsubsection{Static Analysis}
Static analysis tools take the source code as input, compile it, and analyse it for vulnerabilities, errors, and potential optimisations.
Some static analysis tools, such as Slither\cite{feist_slither_2019}, create an Intermediate Representation (IR) of the code to aid in various analysis components. Static analysis tools tend to be relatively fast but often produce false positives. Examples include Slither\cite{feist_slither_2019} and SmartCheck\cite{tikhomirov_smartcheck_2018}.

\subsubsection{Dynamic Analysis}
Dynamic analysis tools analyse the code through execution. They often use a mix of techniques to improve results measured by either time-to-execute or accuracy and fall into two primary categories -
\begin{enumerate}
    \item \textbf{Symbolic Execution - } inputs are treated as symbols and the paths through the program are calculated via constraints using a solver (such as Z3\cite{de_moura_z3_2008}). Examples include Mythril\cite{consensys_mythril_2023}, Oyente\cite{luu_making_2016} and Osiris\cite{torres_osiris_2018}.
    \item \textbf{Fuzzing - } inputs are mutated through iteration and repeated to find unexpected outcomes. Success can be measured by instruction coverage, vulnerabilities detected, and invariants (states set by the user that should not be reachable). Examples of fuzzers are ItyFuzz\cite{shou_ityfuzz_2023}, RLF\cite{su_effectively_2023} and Echidna\cite{grieco_echidna_2020}.
\end{enumerate}

\subsection{LLM usage in Vulnerability Detection}
Before the release of OpenAI's ChatGPT\cite{openai_introducing_2022}, there was already active research into using language models such as GPT-2\cite{radford2019language}, BERT\cite{devlin_bert_2019} and CODEBert\cite{feng_codebert_2020} for solidity code analysis and vulnerability detection. For example, Zeng et al's SolGPT\cite{zeng_solgpt_2024} uses GPT-2\cite{radford2019language} small model for training; Sun et al's Assert\cite{sun_assert_2023}  and Xu et al's SolBERT-BiGRU\cite{xu_vulnerability_2023} use BERT\cite{devlin_bert_2019}. 

\subsection{Literature Scope and Search Parameters}
As the goal of this paper is primarily to investigate the usage of  Generative Large Language Models, we have limited the scope of our search primarily to results from 2021/22 onwards. Also, we have focused exclusively on Ethereum, as Ethereum is currently the most popular and researched smart contract blockchain. The primary search terms were \textit{Ethereum}, \textit{LLMs}, with usage of \textit{vulnerability} when search result refinement was required.

The platforms used were IEEE Explore, ACM Digital Library, Google Scholar,
Springer Link, Web of Science, DBLP Bibliography, and EI Compendex - with most results from Google Scholar or IEEE Explore. We also added additional papers found after reviewing the identified papers.

From the paper search results, we chose 65 relevant papers for deeper analysis (primarily from the years 2023 and 2024). We then investigated the paper in more depth to identify which meet the following criteria;
\begin{enumerate}
    \item The proposed technique focused on vulnerability detection using Generative LLMs
    \item The proposed technique is reproducible (either through prompts or code)
\end{enumerate}

Papers that met the first criteria were included in our study; however, only papers that met both criteria were included in our evaluation. We ended our literature search process with 22 papers for study and 9 for evaluation.

\section{Generative LLM Detection Approaches}
An overview of our review that focuses on the models, techniques, and the training or embedding data that is used can be seen in \cref{tab:llm_usage}.

As shown in \cref{tab:llm_usage}, researchers often use multiple techniques in their proposed tools. This section details the specialisation techniques used and provides examples of their usage in the surveyed papers.

\begin{table*}[htbp]
\centering
\caption{LLM Usage in Smart Contract Analysis Tools Surveyed}
\label{tab:llm_usage}
\begin{tabularx}{\textwidth}{|p{2.5cm}|X|X|X|}
\hline
\textbf{Tool} & \textbf{Base Model} & \textbf{Specialisation Techniques} & \textbf{Training / Embed. Data} \\ \hline
SolGPT \cite{zeng_solgpt_2024} & GPT2 java small \cite{radford2019language} & Specialised Pre-training, custom tokenizer, Supervised Fine-Tuning & \cite{qian_messi-qsmart-contract-dataset_2024} and data crawled from Etherscan \\ \hline
Fine-tuned Llama 2\cite{yang_automated_2024} & LLaMA 2 13B\cite{touvron_llama_2023}, CodeLLaMA 13B\cite{roziere_code_2023} & Supervised Fine-Tuning, PEFT & Labeled functions extracted from Certik audit reports\footnote{Reports were from the period of June 2018 to August 2023 (excluding pen testing reports, duplicate audit reports, and non-solidity reports)} \\ \hline
AuditGPT\cite{xia_auditgpt_2024} & GPT-4-Turbo & In-Context Learning & ERC{20,721,1155} documents summarised \\ \hline
LLM4Vuln\cite{sun_llm4vuln_2024} & None & Evaluation framework & Code4Rena\cite{code_423n4_code_2024} audit reports and findings \\ \hline
LLM4Fuzz\cite{shou_llm4fuzz_2024} & LLaMA 2 70B \cite{touvron_llama_2023} & LLM guided fuzzing and prioritisation & None \\ \hline
ContractArmor \cite{ozdemir_sonmez_contractarmor_2024} & GPT-3 & Contract attack surface analysis, fine-tuning & None \\ \hline
TrustLLM\cite{ma_combining_2024} & Code LLaMA \{13B,34b\}\cite{roziere_code_2023}, Mixtral 8x7B-Instruct\cite{jiang_mixtral_2024}, GPT-4 & Adversarial audit analysis, Supervised Fine-Tuning & Solodit.xyz\cite{solodit_solodit_nodate}, LLM4Vuln\cite{sun_llm4vuln_2024}, GPTScan\cite{sun_when_2023} \\ \hline
PropertyGPT\cite{liu_propertygpt_2024} & GPT-4-Turbo & In-Context Learning, Property Specification Language, Property Generation & None \\ \hline
VulnHunt-GPT\cite{boi_vulnhunt-gpt_2024} & GPT-3.5-Turbo & Prompt engineering, vulnerability description embeddings & OWASP Smart Contract Top 10\cite{varghese_behanan_owasp_2023} \\ \hline
PSCVFinder \cite{yu_pscvfinder_2023} & CodeT5\cite{wang_codet5_2021} & CSCV code slicing, Prompt-tuning, Supervised Fine-Tuning, pre-training & SmartBugs Wild Dataset\cite{ferreira_smartbugs_2020}, ESC dataset\cite{liu_smart_2021} \\ \hline
GPTScan \cite{sun_when_2023} & GPT-\{3.5,4\} & Scenario and property specification, logic vulnerability detection & None \\ \hline
GPTLens \cite{hu_large_2023} & GPT-\{3.5,4\} & Adversarial audit analysis, ranking & None \\ \hline
David et al. \cite{david_you_2023} & GPT-4, Claude & Prompt-tuning, evaluation & None \\ \hline
Detect LLaMA \cite{ince_detect_2024} & Code LLaMA 34B, Code LLaMA 34B Instruct\cite{roziere_code_2023} & Supervised Fine-Tuning & ScrawlD Dataset\cite{yashavant_scrawld_2022} \\ \hline
SmartVD Framework\cite{alam_detection_2024} & CodeLlama 7b\cite{roziere_code_2023} & Custom dataset, Supervised Fine-Tuning, Prompt tuning & VulSmart\cite{alam_detection_2024} \\ \hline
Bouafif et al.\cite{bouafif_context-driven_2024} & GPT-4 & CCL Chunking, In-Context Learning & SmartBugs Curated\cite{ferreira_smartbugs_2020}, SolidiFI Benchmark\cite{ghaleb2020effective} \\ \hline
SmartGuard\cite{ding_smartguard_2024} & CodeBERT\cite{feng_codebert_2020}, GPT-3.5-turbo & CoT generation, In-Context Learning & Messi-Q/Smart-Contract-Dataset (resource 2)\cite{10.1145/3543507.3583367} \\ \hline
FELLMVP\cite{luo_fellmvp_2024} & Gemma 7b\cite{team_gemma_2024} with LlamaFactory\cite{zheng2024llamafactory} & LLM ensemble Agent, Supervised Fine-Tuning & Messi-Q/Smart-Contract-Dataset (resource 3)\cite{liu_rethinking_2023} \\ \hline
LLMSmartSec\cite{mothukuri_llmsmartsec_2024} & GPT-4 & In-Context Learning, Multi-agent analysis & DappScan\cite{zheng_dappscan_2024}, Slither Audit Set\cite{rossini2022slitherauditedcontracts}, EthTrust Security Levels Specification\footnote{https://entethalliance.org/specs/ethtrust-sl/v2/} \\ \hline
FTAudit\cite{wei_leveraging_2024} & Llama 7b\cite{dubey_llama_2024}, Gemma-7b\cite{team_gemma_2024}, CodeGemma 7b\cite{team_codegemma_2024}, Mistral 7b\cite{jiang_mistral_2023} & Knowledge distillation, Multi-agent analysis, Supervised Fine-Tuning & DASP\cite{wong_dasp_2018}, SWC\footnote{SWC Registry - https://swcregistry.io/}, DefiVulnLabs\footnote{https://github.com/SunWeb3Sec/DeFiVulnLabs}, Web3Bugs\cite{DBLP:conf/icse/ZhangZXL23}, Generated synthetic data\cite{wei_leveraging_2024}  \\ \hline
LLM-SmartAudit\cite{wei_llm-smartaudit_2024} & GPT-4o-mini, GPT-3.5-Turbo & Multi-agent analysis & None \\ \hline
ABAuditor\cite{zhang_detecting_2024} & GPT-3.5-Turbo & Prompt-tuning, Rule-based reasoning, hallucination identification, remediation & None \\ \hline
\end{tabularx}
\end{table*}

\subsection{Prompt-tuning}
Prompt-tuning, or prompt engineering, is where specific techniques are used to ensure you get the most accurate or desired results from the large language model. Some examples of this are \textit{Chain of Thought}\cite{wei_chain--thought_2022}, \textit{Few-shot prompting}\cite{wang_generalizing_2020} and \textit{In-Context Learning}\cite{shin_effect_2022,zhu_can_2024}.

Prompt-tuning is often used in conjunction with other techniques. For instance, Boi et al. use a combination of prompt-engineering and context embedding (uses embeddings of the OWASP Smart Contract Top 10\cite{varghese_behanan_owasp_2023}) to assist GPT3.5 Turbo to identify vulnerabilities and provide remediation recommendations\cite{boi_vulnhunt-gpt_2024}.

PropertyGPT\cite{liu_propertygpt_2024} uses retrieval-augmented generation with in-context learning to assist the LLM with the generation of properties from smart contracts for formal verification using their custom Property Specification Language(PSL).

Sun and Wu et al. break down 10 logic vulnerabilities into scenario and property components, which are then used in the prompt for identified candidate functions\cite{sun_when_2023}. By using LLMs in conjunction with static analysis, they reduce the number of false positives output by the LLM, while benefiting from the capacity of the LLM to identify and understand the variables and how they are being used\cite{sun_when_2023}.

Bouafif et al.\cite{bouafif_context-driven_2024} take a slightly different approach in ICL by creating a Code-Call List (a list transformed from the contract function call graph including the codes and path function calls), which is provided to the LLM in conjunction with exploit details and shows that this returns better results than including only the flattened code.

SmartGuard\cite{ding_smartguard_2024} provides an excellent example of novel Chain of Thought construction and validation using a labeled corpus of existing smart contract code that is matched using an LLM to identify the 3 most similar examples. This is then parsed through an iterative self-check CoT process before the final prompt set is sent to the generative LLM for analysis\cite{ding_smartguard_2024}.

\subsection{Hallucination Reduction}
One of the challenges of chaining outputs from LLMs, or utilising LLMs in validation and software products is the potential for hallucinations to go unnoticed. ABAuditor\cite{zhang_detecting_2024} reduces the potential for hallucinations in chains of LLM interactions by interjecting the reflection of prior actions and decisions with the definition of financial terms and appropriate rules.

\subsection{Supervised Fine-tuning}
Fine-tuning a model involves taking a pre-trained model and specialising it for your specific purpose, domain or tasks. Whereas pre-training an LLM is typically unsupervised on a corpus, fine-tuning (specifically generative LLMs such as GPT-3.5 Turbo\cite{peng_gpt-35_2023} and Meta's Code Llama models\cite{roziere_code_2023}),  often uses a prompt-template (such as Alpaca Instruct\cite{taori_alpaca_nodate}).

In \cite{ince_detect_2024}, Ince et al fine-tune two Code Llama 34b models\cite{roziere_code_2023} using two primary prompt styles, generation and detection, with entire smart contracts with labels (excluding comments and extra lines). \cite{ince_detect_2024} utilises techniques such as Flash Attention 2\cite{dao_flashattention-2_2023} and QLoRA\cite{dettmers_qlora_2023} to reduce the hardware requirements to train such a large model.

Yang et al. fine-tune Code-Llama and Llama 2 13b parameter models on function level vulnerability detection - evaluating their results against standard Code-Llama and Llama 2 13b models\cite{yang_automated_2024}. 

PSCVFinder\cite{yu_pscvfinder_2023} usea a different approach to fine-tuning; the smart contracts are processed using a novel CSCV (Crucial Smart Contract for Vulnerabilities) representation (both in the labelled dataset and for processing). This processing normalises the variable and function names, and removes part of the code that does not meet the following criteria;
\begin{itemize}
    \item Statements containing code directly related to the vulnerability
    \item Data-dependent statements
    \item Control-dependent statements
\end{itemize}

Utilising the CSCV representation in combination with other techniques in \cite{yu_pscvfinder_2023},  PSCVFinder can out-perform the static and dynamic analysis tools they chose for baseline in detection of Reentrancy and Timestamp dependence vulnerabilities\cite{yu_pscvfinder_2023}. 

ContractArmor\cite{ozdemir_sonmez_contractarmor_2024} uses fine-tuning as a method to improve poor performance on specific sets of questions from the attack surface generator.

Supervised Fine-tuning's effectiveness is multiplied by the quality of the data used. SmartVD Framework\cite{alam_detection_2024} creates a custom dataset that is \textit{balanced} (i.e. - has an equal number of examples for each vulnerability) across 13 different vulnerability types, ensuring each targeted vulnerability receives the same amount of fine-tuning inputs. 

\subsection{Ensemble LLMs}
FELLMVP\cite{luo_fellmvp_2024} shows a unique and innovative approach of utilising Supervised Fine-tuning on 1 smaller (7b parameter) model per vulnerability, reducing the complexity requirements for each instruction as the output is binary as to whether the specific vulnerabilitiy is identified.

\subsection{Model Pre-training}
Pre-training is a process of unsupervised learning that is performed on a corpus of data. In LLMs like OpenAI's GPT-4\cite{openai_gpt-4_2023}  and Meta's Llama models\cite{touvron_llama_2023,roziere_code_2023}, the corpus is typically internet scale - a huge amount of data sourced from crawling the internet, social sites, and open-source code repositories.

However, pre-training can also be more targeted. SolGPT\cite{zeng_solgpt_2024} uses a targeted dataset of 726 samples to further pre-train the GPT-2 java small model\cite{radford2019language} on the unlabelled smart contract function data, before fine-tuning the model on the same dataset with vulnerability detection labels attached. SolGPT also uses a custom tokenizer, \textit{SolTokenizer}, that utilises the Byte-Pair Encoding (BPE) algorithm\cite{sennrich_neural_2016} to produce improved results for Solidity syntax tokenisation in pre-training and fine-tuning processes\cite{zeng_solgpt_2024}.

\subsection{Dynamic Guiding}
In fuzzing, the number of potential transaction combinations and mutations often makes testing parts of the smart contract more time-consuming. One approach to reducing these constraints is to use some form of \textit{guiding} - a technique, or combination of techniques, to aid the fuzzer in prioritising mutation combinations or instructions to improve efficiency.

LLM4Fuzz\cite{shou_llm4fuzz_2024} uses the Llama 2 70b model\cite{touvron_llama_2023} to measure complexity, vulnerability likelihood, sequential likelihood and other measures to prioritise and guide scheduler for fuzzing targets. This technique can identify previously unknown vulnerabilities and outperform a current State-of-the-art fuzzer, ItyFuzz\cite{shou_ityfuzz_2023}\cite{shou_llm4fuzz_2024}.

\subsection{Multi-Agent Analysis}
When discussing agents within the context of LLMs, an agent is typically an instance of an LLM that is given a prompt to behave or act in a specific role, sometimes given a specific perception and expected output.

In LLMSmartSec\cite{mothukuri_llmsmartsec_2024}, three agents are used with a vector store of relevant information to provide individual analysis summarised to produce the final report. These three agents are; \textit{LLMeHack} - a smart contract hacker to identify and provide details of valuable real-world exploits, \textit{LLMDev} - a smart contract developer to analyse as a developer, and \textit{LLMAudit} - a smart contract auditor to provide a detailed report of any risks or vulnerabilities\cite{mothukuri_llmsmartsec_2024}.

Another approach to the multi-agent analysis is to have the agents work together with different roles. LLM-SmartAudit\cite{wei_llm-smartaudit_2024} uses a multi-agent approach that specifies a set number of agents with different roles and has them collaborate to identify vulnerabilities and provide an output. LLM-SmartAudit proposes that 5-6 agents examine the smart contract through their individual roles while working toward an overarching team collaborative goal\cite{wei_llm-smartaudit_2024}.

FTAudit\cite{wei_leveraging_2024} uses multi-agent analysis in a different part of the process - a Distillation agent, a Developer agent and a Security agent are used to take the records from the selected datasets and transform them into synthetic data following a provided template and structure. This transformed data is then used for Supervised fine-tuning of their model\cite{wei_leveraging_2024}. 

\subsection{Adversarial Analysis}
In adversarial analysis, two (or more) agents perform an $analysis\Rightarrow critique \Rightarrow rank$ process, allowing for improvement and refinement of vulnerability detection. By adding a critic agent to $n$ auditors, \cite{hu_large_2023} shows they can achieve better accuracy for vulnerability detection.  

TrustLLM\cite{ma_combining_2024} took the adversarial agent analysis a step further. Four agents are used - the \textit{Detector} and \textit{Reasoner} agents are each specifically fine-tuned using LoRA\cite{hu_lora_2021} for their specific tasks. The two other agents are based on Mistral's Mixtral8x7b Instruct model\cite{jiang_mixtral_2024} to act as \textit{Ranker} and \textit{Crtic}\cite{ma_combining_2024}.

\subsection{Evaluation}
\cite{david_you_2023} was one of the first evaluations on the use of generative large language models GPT-4-32k\cite{openai_gpt-4_2023} and Claude v1.3-100k as smart contract vulnerability detectors through prompt only\cite{david_you_2023}.
In \cite{david_you_2023}, 52 DeFi projects that had previously been attacked are analysed, and each prompt provides the smart contract, the vulnerability to detect, and how the model should respond. 

LLM4Vuln is a comprehensive framework for evaluating different large language models as smart contract vulnerability detectors\cite{sun_llm4vuln_2024}. LLM4Vuln aims to separate the LLMs reasoning ability from the other abilities and measure their capability with and without tools such as knowledge retrieval (e.g., RAG), tool invocation (e.g., function calling), prompt schemes (e.g., Chain of Thought) and instruction following\cite{sun_llm4vuln_2024}.

\section{LLM-based Tool Evaluation}\label{sec:eval}

\subsection{Tool selection}
At the time of evaluation, only seven of the analysed tools had their code (or model when required) open-sourced: PSCVFinder\cite{yu_pscvfinder_2023}, GPTScan\cite{sun_when_2023}, GPTLens\cite{hu_large_2023}, Detect Llama\cite{ince_detect_2024}, LLM-SmartAudit\cite{wei_llm-smartaudit_2024}, FTAudit\cite{wei_leveraging_2024}, LLMSmartSec\cite{mothukuri_llmsmartsec_2024} and Bouafif et al.\cite{bouafif_context-driven_2024}. 

Also, while \cite{david_you_2023}'s evaluation of GPT-4 and Claude was prompt-based and did not include any further tool, the prompts in the paper can be replicated.

The other papers generally fall into three categories;
\begin{enumerate}
    \item Papers that made no mention of release. This includes AuditGPT\cite{xia_auditgpt_2024}, VulntHunt-GPT\cite{boi_vulnhunt-gpt_2024}, ContractArmor\cite{ozdemir_sonmez_contractarmor_2024}, Yang and Man et al's work\cite{yang_automated_2024}, LLM4Vuln\cite{sun_llm4vuln_2024}, SolGPT\cite{zeng_solgpt_2024}, SmartGuard\cite{ding_smartguard_2024}, FELLMVP\cite{luo_fellmvp_2024} and ABAuditor\cite{zhang_detecting_2024}.
    \item Papers that mention (or link to a mention) of their tool being made available post-paper acceptance or sometime in the future - in some cases, they provide data. This includes PropertyGPT\cite{liu_propertygpt_2024}, LLM4Fuzz\cite{shou_llm4fuzz_2024} and \cite{alam_detection_2024}.
    \item Papers that chose not to release their tool for ethical concerns around financial risk in DeFi. This includes \cite{ma_combining_2024}.
\end{enumerate}

Unfortunately, four of the open-source tools were not able to be included in our evaluation; 
PSCVFinder\cite{yu_pscvfinder_2023} seemed to be missing a component and could not be run by our evaluator, and we could not get further information from the corresponding author of the paper.
GPTScan\cite{sun_when_2023} was evaluated as a tool; however, the program issues it evaluates for did not match our dataset or other tools being evaluated.
\cite{bouafif_context-driven_2024} is designed more for individual co-auditing and not as an automated process\footnote{The tool uses headless selenium testing vs OpenAI's API}.
LLMSmartSec\cite{mothukuri_llmsmartsec_2024} is missing information on how the OpenAI agents and setup and used with the prompts.

We also evaluate against Slither\cite{feist_slither_2019} and Mythril\cite{consensys_mythril_2023} to view how LLM-based tools compare against more traditional static and dynamic analysis tools.

\subsection{Dataset selection}
W wanted to use a dataset that minimized data contamination by not being used by any models being evaluated.

We selected the dataset \textit{Vulnerable verified smart contracts}\cite{Storhaug2023} by Storhaug. The dataset contains 609 vulnerable contracts, containing 1,117 vulnerabilities over ten distinct vulnerability types\cite{Storhaug2023}. The dataset was developed for Storhaug et al's paper\cite{storhaug_efficient_2023} and focused on the vulnerability types identified as: \textit{DelegateCall}, \textit{Nested Call}, \textit{Reentrancy}, \textit{Timestamp Dependency}, \textit{Transaction Order Dependency}, \textit{Unchecked Call}, \textit{Unprotected Suicide} and \textit{Frozen Ether}. This dataset meets our criteria as it was not used to train the tools we evaluated.

Another criterion for dataset choice is the vulnerabilities detected by the tools to be evaluated. The tools that we selected either did not specify the vulnerabilities for detection, or have vulnerabilities targeted that largely fit within the dataset. For example, Detect Llama\cite{ince_detect_2024} uses eight pre-specified vulnerabilities, 7 of which are matched by the test dataset. For instances where the model does not identify a specific vulnerability in it's design, we label the results as \textit{N/A}.

\subsection{Environment Setup}
GPTLens\cite{hu_large_2023}, David et al.'s prompts\cite{david_you_2023}, LLMSmartAudit\cite{wei_llm-smartaudit_2024}, Mythril\cite{consensys_mythril_2023} and Slither\cite{feist_slither_2019} were all run on an Intel NUC device with an eight-core 11th Gen Intel i7  at 4.7GHz and 64GB of RAM. Tools that require GPU - Detect Llama\cite{ince_detect_2024} and FTAudit\cite{wei_leveraging_2024} - were run on a runpod.io\footnote{Runpod.io provide relatively cheap on-demand GPU images - https://www.runpod.io/} container using a modified huggingface text-generation-inference image and 1 A100 SXM GPU with 80GB VRAM.

\subsection{Evaluation Results}

The results from our comparative evaluation can be seen in \cref{tab:perf}; this section includes a model summary, vulnerability description and analysis of the results.

\subsubsection{Model summary}
\begin{itemize}
    \item \textbf{the prompts from David et al.} (referred to as David et al.)\cite{david_you_2023} had each vulnerability and their description added to the prompt, as per the paper, and was executed against each contract once per vulnerability. 
    \item \textbf{GPTLens}\cite{hu_large_2023} processed each contract once by the auditor function and separately by the critic function. We manually matched the results as the standard prompt for GPTLens does not specify which vulnerabilities to look for.
    \item \textbf{GPTLens def.} is \cite{hu_large_2023}, but we have added a list of the vulnerabilities being sought to the audit prompt.
    \item \textbf{Detect Llama}\cite{ince_detect_2024} was executed once per contract without any modifications
    \item \textbf{FTAudit}\cite{wei_leveraging_2024} processed each contract once performed post-processing on the responses to match the identified vulnerabilities to the dataset
    \item \textbf{FTAudit def.} is \cite{wei_leveraging_2024} with the details of the vulnerabilities to look for added to the prompt
    \item \textbf{LLM SmartAudit}\cite{wei_llm-smartaudit_2024} processes each smart contract using SmartAudit\_TA and processes the response as a report with the targeted vulnerabilities
    \item \textbf{Mythril}\cite{consensys_mythril_2023} was executed once per contract with an execution timeout added of 300 seconds
    \item \textbf{Slither}\cite{feist_slither_2019} was executed once per contract without any modifications
\end{itemize}
23 smart contracts were excluded from the analysis performed using language models as, even after removing comments, their length was beyond 7500 tokens (a token is, on average, four letters).

\subsubsection{Vulnerabilities}
We have included 8 of the 10 vulnerabilities from the dataset\cite{Storhaug2023}; \textbf{DelegateCall (DC)}, \textbf{Frozen Ether (FE)}, \textbf{Integer Overflow/Underflow (IO)}, \textbf{Reentrancy (RE)}, \textbf{Timestamp Dependency (TD)}, \textbf{Transaction Order Dependency (TOD)}, \textbf{TxOrigin (TO)},\textbf{Unchecked Call (UC)}.
\begin{table*}[!ht]
\centering
\caption{Performance Metrics for Evaluated Tools}
\label{tab:perf}
\begin{tabular}{|l|l|c|c|c|c|c|c|c|c|}
\hline
\textbf{Model}     & \textbf{Measure} & \textbf{DC} & \textbf{FE} & \textbf{IO} & \textbf{RE} & \textbf{TD} & \textbf{TOD} & \textbf{TO} & \textbf{UC} \\ \hline

David et al. & Accuracy   & 0.96 & 0.25 & 0.57 & 0.4  & 0.84 & 0.13 & 0.72 & 0.26 \\ \hline
\rowcolor{yellow} \textbf{David et al.} & \textbf{F1 Score} & \textbf{0.79} & \textbf{0.01} & \textbf{0.37} & \textbf{0.27} & \textbf{0.77} & \textbf{0.00} & \textbf{0.15} & \textbf{0.12} \\ \hline
David et al. & Precision  & 0.73 & 0.00 & 0.28 & 0.16 & 0.65 & 0.00 & 0.08 & 0.07 \\ \hline
David et al. & Recall     & 0.86 & 1.00 & 0.55 & 1.00 & 0.95 & 0.00 & 1.00 & 1.00 \\ \hline

GPTLens      & Accuracy   & 0.96 & 1.00 & 0.75 & 0.62 & 0.73 & 0.74 & 0.98 & 0.94 \\ \hline
\rowcolor{yellow} \textbf{GPTLens} & \textbf{F1 Score} & \textbf{0.66} & \textbf{0.00} & \textbf{0.04} & \textbf{0.34} & \textbf{0.01} & \textbf{0.00} & \textbf{0.46} & \textbf{0.05} \\ \hline
GPTLens      & Precision  & 1.00 & 0.00 & 0.21 & 0.20 & 0.50 & 0.00 & 0.55 & 0.12 \\ \hline
GPTLens      & Recall     & 0.49 & 0.00 & 0.02 & 1.00 & 0.01 & 0.00 & 0.40 & 0.03 \\ \hline

GPTLens def  & Accuracy   & 0.98 & 0.88 & 0.76 & 0.54 & 0.77 & 0.73 & 0.86 & 0.69 \\ \hline
\rowcolor{yellow} \textbf{GPTLens def} & \textbf{F1 Score} & \textbf{0.83} & \textbf{0.00} & \textbf{0.12} & \textbf{0.30} & \textbf{0.42} & \textbf{0.00} & \textbf{0.24} & \textbf{0.19} \\ \hline
GPTLens def  & Precision  & 0.90 & 0.00 & 0.37 & 0.17 & 0.67 & 0.00 & 0.14 & 0.11 \\ \hline
GPTLens def  & Recall     & 0.78 & 0.00 & 0.07 & 1.00 & 0.31 & 0.00 & 0.93 & 0.77 \\ \hline

Detect Llama & Accuracy   & 0.94 & 1.00 & 0.32 & 0.79 & 0.73 & 0.79 & 0.98 & 0.95 \\ \hline
\rowcolor{yellow} \textbf{Detect Llama} & \textbf{F1 Score} & \textbf{N/A} & \textbf{0.00} & \textbf{0.41} & \textbf{0.00} & \textbf{0.16} & \textbf{0.70} & \textbf{0.00} & \textbf{0.20} \\ \hline
Detect Llama & Precision  & N/A  & 0.00 & 0.26 & 0.00 & 0.62 & 0.58 & 0.00 & 0.60 \\ \hline
Detect Llama & Recall     & N/A  & 0.00 & 0.95 & 0.00 & 0.09 & 0.87 & 0.00 & 0.12 \\ \hline

FTAudit      & Accuracy   & N/A  & 1.00 & 0.53 & 0.13 & 0.76 & 0.81 & 0.98 & 0.94 \\ \hline
\rowcolor{yellow} \textbf{FTAudit} & \textbf{F1 Score} & \textbf{0.34} & \textbf{0.00} & \textbf{0.38} & \textbf{0.20} & \textbf{0.23} & \textbf{0.00} & \textbf{0.00} & \textbf{0.06} \\ \hline
FTAudit      & Precision  & 0.64 & 0.00 & 0.26 & 0.11 & 0.72 & 0.00 & 0.00 & 0.08 \\ \hline
FTAudit      & Recall     & 0.23 & 0.00 & 0.70 & 1.00 & 0.13 & 0.00 & 0.00 & 0.05 \\ \hline

FTAudit def  & Accuracy   & 0.92 & 0.01 & 0.19 & 0.13 & 0.27 & 0.20 & 0.03 & 0.04 \\ \hline
\rowcolor{yellow} \textbf{FTAudit def} & \textbf{F1 Score} & \textbf{0.17} & \textbf{0.00} & \textbf{0.32} & \textbf{0.22} & \textbf{0.42} & \textbf{0.32} & \textbf{0.03} & \textbf{0.08} \\ \hline
FTAudit def  & Precision  & 0.17 & 0.00 & 0.19 & 0.12 & 0.27 & 0.19 & 0.02 & 0.04 \\ \hline
FTAudit def  & Recall     & 0.17 & 1.00 & 1.00 & 1.00 & 0.99 & 0.93 & 1.00 & 1.00 \\ \hline

LLM SmartAudit & Accuracy   & N/A  & N/A  & 0.58 & 0.87 & N/A  & 0.46 & 0.98 & 0.41 \\ \hline
\rowcolor{yellow} \textbf{LLM SmartAudit} & \textbf{F1 Score} & \textbf{N/A} & \textbf{N/A} & \textbf{0.52} & \textbf{0.00} & \textbf{N/A} & \textbf{0.47} & \textbf{0.14} & \textbf{0.06} \\ \hline
LLM SmartAudit & Precision  & N/A  & N/A  & 0.36 & 0.00 & N/A  & 0.32 & 0.33 & 0.03 \\ \hline
LLM SmartAudit & Recall     & N/A  & N/A  & 0.90 & 0.00 & N/A  & 0.92 & 0.09 & 0.33 \\ \hline

mythril       & Accuracy   & 0.94 & 1.00 & 0.46 & 0.85 & 0.92 & 0.67 & 0.80 & 0.97 \\ \hline
\rowcolor{yellow} \textbf{mythril} & \textbf{F1 Score} & \textbf{0.38} & \textbf{0.00} & \textbf{0.26} & \textbf{0.59} & \textbf{0.86} & \textbf{0.14} & \textbf{0.17} & \textbf{0.62} \\ \hline
mythril       & Precision  & 1.00 & 0.00 & 0.19 & 0.43 & 0.85 & 0.20 & 0.10 & 0.82 \\ \hline
mythril       & Recall     & 0.23 & 0.00 & 0.40 & 0.93 & 0.87 & 0.10 & 0.85 & 0.50 \\ \hline

slither       & Accuracy   & 0.95 & 1.00 & 0.77 & 0.91 & 0.87 & 0.75 & 0.99 & 0.97 \\ \hline
\rowcolor{yellow} \textbf{slither} & \textbf{F1 Score} & \textbf{0.54} & \textbf{0.50} & \textbf{0.00} & \textbf{0.68} & \textbf{0.78} & \textbf{0.00} & \textbf{0.75} & \textbf{0.57} \\ \hline
slither       & Precision  & 1.00 & 0.33 & 0.00 & 0.55 & 0.74 & 0.00 & 0.82 & 0.86 \\ \hline
slither       & Recall     & 0.37 & 1.00 & 0.00 & 0.86 & 0.83 & 0.00 & 0.69 & 0.43 \\ \hline

\end{tabular}
\end{table*}
\subsubsection{Analysis}

\cref{tab:perf} evaluates the correctness of the results generated by the tools using the following metrics;
\begin{itemize}
    \item \textbf{Accuracy} measures how many predicted values matched the actual values 
    \[\frac{TP+TN}{TP+TN+FP+FN}\]
    \item \textbf{Precision} measures the ratio of correctly predicted positive values vs all predicted positive values
    \[\frac{TP}{TP+FP}\]
    \item \textbf{Recall} measures the ratio of correctly predicted positive values vs all predicted values
    \[\frac{TP}{TP+FN}\]
    \item \textbf{F1 Score} can be referred to as the harmonic mean of \textit{Precision} and \textit{Recall}, provides a good overall score of the model
    \[\frac{2 \times (Precision \times Recall)}{Precision + Recall}\]
\end{itemize}

We can see that generally, the non-LLM-based tools perform better on average, but the LLM-based tools perform better on some vulnerabilities.
For instance, for the \textit{DelegateCall} vulnerability David et al., GPTLens and GPTLens def. all outperform Mythril and Slither with F1 Scores of $0.79$, $0.66$ and $0.83$ for the LLM tools respectively, compared to $0.38$ and $0.54$ for Mythril and Slither.

In summary, the traditional tools performed significantly better at detecting Frozen Ether, Reentrancy, and Unchecked Call vulnerabilities; LLM tools performed better at detecting Transaction Order Dependency, Integer Overflow/Underflow, and Delegate Call; and results are mixed for Tx.Origin, and Timestamp Dependency.

\subsubsection{Difference in results} In Detect Llama\cite{ince_detect_2024}, their model is compared against \cite{hu_large_2023} using the same split method. However, \cite{ince_detect_2024}'s Foundation model significantly outperforms the GPTLens technique and the GPTLens def. variant\cite{ince_detect_2024} whereas our results in \cref{tab:perf} find that the model outperforms GPTLens; however, it performs similarly when compared to the GPTLens def. variant\footnote{excluding GPTLens def.'s excellent performance in detection of \textit{DelegateCall}, as Detect Llama does not support detection of this vulnerability}. The performance difference is likely due to Detect Llama being fit specifically onto the dataset/process used (\cite{yashavant_scrawld_2022}). It is also possible that because we utilised GPT-4o for the model supporting GPTLens, the small improvements in the benchmarks\cite{openai_hello_2024} represented an improvement in GPTLens def. However, the data in \cref{tab:perf} indicates that Detect Llama performed worse than in \cite{ince_detect_2024}.

\subsection{Performance}

As shown in  \cref{tab:runtime}, the timing varies significantly for traditional and LLM-based tools. Analysing the non-LLM-based tools, the results are similar to what we would expect - Mythril, the symbolic execution tool, takes much longer than Slither, the static analysis tool.

\begin{table}[!ht]
\centering
\caption{Runtime per analysis in seconds}
\label{tab:runtime}
\begin{tabular}{|l|r|r|r|r|r|}
\hline
\textbf{Tool}       & \textbf{Mean} & \textbf{Median} & \textbf{Std. Dev} & \textbf{Min.} & \textbf{Max.} \\ \hline
GPTLens             & 21.23        & 18.81          & 10.86            & 4.36          & 155.01       \\ \hline
GPTLens def.        & 14.1         & 12.7           & 6.08             & 4.45          & 73.7         \\ \hline
Detect llama        & 2.8          & 2.33           & 1.8              & 1.3           & 8.23         \\ \hline
David et al.        & 7.38         & 6.47           & 3.01             & 5.05          & 36.24        \\ \hline
FTAudit             & 105.66       & 112.73         & 22.73            & 20.07         & 116.85       \\ \hline
FTAudit def         & 81.2         & 84.72          & 16.99            & 16.01         & 103.14       \\ \hline
LLMSmartAudit       & 127.15       & 124.73         & 38.14            & 59.73         & 738.74       \\ \hline
Mythril             & 430.25       & 313.49         & 466.87           & 2.5           & 2985.62      \\ \hline
Slither             & 0.57         & 0.42           & 0.33             & 0.36          & 3.56         \\ \hline
\end{tabular}
\end{table}

The two GPTLens tools have two processes we measure for time and token usage: audit and critic\cite{hu_large_2023}.  However, for comparison, we have combined them in \cref{tab:runtime,tab:token_usage}. 
Detect Llama was the fastest of the LLM-based tools, followed by David et al. and GPTLens.

The results show that the two FTAudit-based analyses took longer than the Detect Llama analyses. This is because the FTAudit model is trained to provide more comprehensive descriptions of the vulnerabilities found and a description of the problem (typically 1-2 thousand tokens), whereas Detect Llama was specifically trained to only reply with the names of the vulnerabilities found—and it is the generation of tokens that is the time-intensive part of using an LLM.

When the results from \cref{tab:runtime} and \cref{tab:token_usage} are viewed together, we can see that generating a larger amount of tokens strongly indicates how long the tool takes to return its results. However, the exception to this is \cite{david_you_2023}; this is due to the performance of a full analysis per contract and vulnerability, with the results being \textit{YES} or \textit{NO} only. The 4872 tokens for David et al. were generated over 4872 API calls, which added processing time.

One of the outliers in terms of context provided and tokens generated is LLMSmartAudit - this is due to the novel conversational and collaborative approach taken with the agents - as each agent must provide the output of the other agents for the reflection process. It is noteworthy that although LLMSmartAudit uses an order of magnitude more than most of the other tools, it still has the second cheapest run-cost - this is due to their use of GPT-3.5-Turbo and GPT-4o-mini for the token-intensive tasks.

\begin{table}[htbp]
\centering
\caption{Token Usage and Costs for Evaluated Tools}
\label{tab:token_usage}
\begin{tabular}{|l|r|r|r|}
\hline
\textbf{Tool}        & \textbf{Context Tokens} & \textbf{Gen. Tokens} & \textbf{Cost (USD)} \\ \hline
GPTLens              & 1,236,016           & 681,880          & 16.49              \\ \hline
GPTLens def.         & 1,263,630           & 669,040          & 16.35              \\ \hline
Detect llama         & 501,385             & 21,037           & 2.50               \\ \hline
David et al.         & 10,664,761          & 4,872            & 53.39              \\ \hline
LLMSmartAudit        & 70,432,209          & 3,639,799        & 9.77               \\ \hline
FTAudit              & 545,959             & 958,168          & 28.00              \\ \hline
FTAudit def          & 733,840             & 881,033          & 22.00              \\ \hline
\end{tabular}
\end{table}

Taking all of the factors shown in \cref{tab:perf,tab:runtime,tab:token_usage} into consideration; none of the evaluated tools performed generally well enough to replace Slither or Mythril, however, their outperformance in some tasks make them a valuable candidate for inclusion in an audit workflow.

\subsubsection{Other Considerations}
While Detect Llama\cite{ince_detect_2024} was faster and cheaper than the other LLM-based tools, the numbers shown do not represent that renting an A100 NVIDIA GPU was required, so while it was cheaper and faster for evaluation, the additional work required is likely prohibitive for small batches of contracts.

\section{Discussion}

The results in \cref{sec:eval} show that none of the LLM-based tools, using fine-tuning or prompting techniques, are ready to replace more traditional static and dynamic analysis tools for vulnerability detection in smart contracts. 

\subsection{Bigger does not mean better}

However, some of the tools not available for evaluation (such as SolGPT\cite{zeng_solgpt_2024} and PSCVFinder\cite{yu_pscvfinder_2023}) did show promise at out-performing more traditional tools in their respective papers. PSCVFinder\cite{yu_pscvfinder_2023} and SolGPT\cite{zeng_solgpt_2024} focus on fewer vulnerabilities (2 and 4, respectively), utilise some form of customised pre-training and focus on smaller context windows.

\subsubsection{PSCVFinder}For instance, in \cite{yu_pscvfinder_2023} Yu et al. utilise a novel normalisation and abstraction process, Crucial Smart Contract for Vulnerabilities (CSCV), in which they gather the required variables that have data, control or other dependence on the code for analysis\cite{yu_pscvfinder_2023}. In addition to contributing to training and detection, the CSCV normalisation and abstraction process aids in fitting the vulnerable functions into the 512 token max input window\cite{yu_pscvfinder_2023}. Yu et al. then continue the pre-training on the base CodeT5 model (200 million parameters)\cite{wang_codet5_2021} with the smart contract detection data.

The PSCVFinder tool, utilising the model (and the normalisation process),  was able to outperform deep learning based methods including \textit{LTSM}, \textit{GRU}, \textit{GCN}, \textit{DR-GCN}, \textit{TMP}, \textit{CGE}, \textit{AME}, \textit{Peculiar}, \textit{ReVulDL}, and \textit{Bi-GGNN} and traditional tools including \textit{Manitcore}, \textit{Mythril}, \textit{Osiris}, \textit{Oyente}, \textit{Slither}, \textit{Securify} and \textit{Smartcheck} on both reentrancy and timestamp dependency vulnerabilities\cite{yu_pscvfinder_2023}.

\subsubsection{SolGPT} In \cite{zeng_solgpt_2024}, Yu et al. develop a specialised tokenizer, \textit{SolTokenizer}, for working with Solidity code, and add a pre-training stage focused on Solidity code, \textit{Solidity Adaptive Pre-training}. Then, a fine-tuning process focuses on four vulnerability types (reentrancy, deletegatecall, timestamp and integer overflow) as a vulnerability detection classification layer\cite{zeng_solgpt_2024}.

Utilising GPT small\cite{radford2019language} as the base model, a 124 million parameter model, the pre-training and fine-tuning process is completed and evaluated against, and out-performs,  existing deep learning approaches including \textit{RNN}, \textit{LSTM}, \textit{BiLSTM}, \textit{BiLSTM-ATT} and \textit{TMP}, and traditional tools including \textit{Slither}, \textit{Mythril} and \textit{Oyente}.

Both SolGPT\cite{zeng_solgpt_2024} and PSCVFinder\cite{yu_pscvfinder_2023} also fine-tuned their models on individual examples per vulnerability instead of multiple vulnerabilities at a time like \cite{ince_detect_2024}.

\subsubsection{Insights}
By utilising pre-training, in conjunction with clever specialisation techniques,  PCSVFinder\cite{yu_pscvfinder_2023} and SolGPT\cite{zeng_solgpt_2024} were both able to out-perform traditional tools in vulnerability detection, and were able to do so using small models. For instance, \cite{yang_automated_2024} tests two types of 13 billion parameter models that are fine-tuned for detection, and \cite{ince_detect_2024} fine-tuned 34 billion parameter models and are not able to achieve accuracy or F1 scores meeting the average in \cite{yu_pscvfinder_2023} or \cite{zeng_solgpt_2024} on a single vulnerability type\cite{ince_detect_2024}. For comparison, 124 million and 200 million parameters in \cite{zeng_solgpt_2024} and \cite{yu_pscvfinder_2023} respectively, vs 13 billion and 34 billion parameters in \cite{yang_automated_2024} and \cite{ince_detect_2024} respectively, making the more accurate and better-performing models 98\% smaller. 

\subsection{The value of larger LLMs as support}

For the larger generative LLMs, such as those from Meta, Anthropic, and OpenAI, the most effective approach seen in our research is to blend static or dynamic analysis tools with LLM support in some form of guidance. Examples of this are LLM4Fuzz\cite{shou_llm4fuzz_2024}, which outperforms unmodified ItyFuzz\cite{shou_ityfuzz_2023} by adding program analysis-based fuzzing guidance to the unmodified tool, and PropertyGPT\cite{liu_propertygpt_2024} which utilises retrieval-augmented generation and static analysis in conjunction with an LLM and their custom Property Specification Language to generate properties for usage in formal verification.


\subsection{Opportunities for future work}
\subsubsection{Ensemble Agents using much smaller models}
FELLMVP\cite{luo_fellmvp_2024} shows excellent promise with their LLM ensemble, however, the compute requirements required for 8 simultaneous Gemma 7b models is quite large. If the pre-training is applied to the GPT-2 small model (as shown in SolGPT\cite{zeng_solgpt_2024}) in addition to the Supervised Fine-tuning performed in \cite{luo_fellmvp_2024}, you could potentially have a similar quality for approximately 1 billion parameters ($8 x 120m = 960m$) vs the original cumulative 56 billion parameters ($8 x 7b = 56b$).

\section{Conclusion}
Our paper presents a detailed and comprehensive study of the use of generative large language models in smart contract vulnerability detection. We analyse their method of action, usage, training data, and specialisation techniques.

We then evaluate 5 of the surveyed tools against Slither and Mythril, and identify the items the traditional tools detected better (Frozen Ether, Reentrancy and Unchecked Call), the items that were mixed (Integer Overflow/Underflow, Tx.Origin and Timestamp Dependency) and items were some of the LLM-based tools outperformed (Transaction Order Dependency, DelegateCall).

The performance of the LLM tools is then analysed, including the wider performance of LLMs against traditional tools and how they can best be used.

For future work, we will use these insights to develop a hybrid tool utilising LLMs to guide a state-of-the-art tool such as ItyFuzz\cite{shou_ityfuzz_2023}.

\bibliographystyle{IEEEtran}
\bibliography{references}

\end{document}